# Contractible Spaces, Homotopy Equivalence and Homeomorphism in Digital Topology

**Alexander V. Evako**

e-mail: evakoa@mail.ru

NPK NOVOTEK, Moscow, Russia

**Abstract** This article provides a brief overview of the main results in the field of contractible digital spaces and contractible transformations of digital spaces and contains new results. We introduce new types of contractible digital spaces such as the cone and the double cone. Based on this, we introduce new contractible transformations that covert the digital space into a homotopy equivalent to the first one. We group together these transformations and get 6 types of contractible transformations. These transformations can be used to convert a closed digital n-dimensional manifold into another closed n-dimensional manifold homeomorphic to the first one.

## 1. Introduction

A digital approach to geometry and topology plays an important role in analyzing n- dimensional digitized images arising in computer graphics as well as in many areas of science including neuroscience, medical imaging, industrial inspection, geoscience and fluid dynamics.

A significant number of works are devoted to the study of the structures of finite sets of points with integer coordinates in Euclidean space with suitable adjacency relation between the points (e.g. [1-5, 17, 32-34, 36]). The results of this approach are used to refine and justify some important low-level image processing algorithms, including algorithms for thinning, boundary extraction, object counting, and contour filling (e.g. [1-5, 17]).

Our approach for constructing digital spaces was introduced and studied in works [6-9, 18-30]. This approach is based on so called molecular spaces, which are families of equal cubes with the edge L in Euclidean space E of dimension 1, 2, 3,…. [18-26, 13].

An important feature of this approach is that it is based on computer experiments and observations. Let G and H be two homeomorphic or homotopy equivalent spaces in a Euclidean space E. We obtain molecular spaces M(G) and M(H), where M(G) is the family of cubes intersecting G and M(H) is the family of cubes intersecting H. Then construct digital models D(G) and D(H) of G and H, which are the intersection graphs of M(G) and M(H). It turned out that if the edge L is small enough, then M(G) (D(G)) can be transformed into M(H) (D(H)) by a sequence of so called contractible transformations. It is reasonable to assume that molecular spaces and digital models contain topological and, perhaps, geometrical characteristics of continuous spaces and contractible transformations digitally model a homeomorphism and a homotopy equivalence of continuous spaces. Thus, molecular spaces and digital models are discrete counterparts of continuous spaces. This approach has a wide range of applications. For example, digital models of some spaces, including the torus, the projective plane and spheres, are presented in [6-7, 9, 31]. Digital models were used to describe a discrete physical space and discrete models of a closed physical Universe [6]. The classification of continuous manifolds of dimension 4, 5 and more by using the classification of their digital models is proposed in [12]. Using digital models, we can numerically solve partial differential equations on manifolds (e.g. [16]) and so on. A number of papers have shown how to construct digital models of continuous spaces using special coverings of these spaces (see, for example, [7-9, 11]). In the present paper we construct new types of contractible digital spaces that are homotopy equivalent to a single point. In particular, we introduce the double cone and prove that the double cone is a contractible digital space. Based on these types of contractible digital spaces, we construct new contractible transformations that allow us to convert a digital space to another digital space homotopy equivalent to the first one, and a digital closed n-manifold to another digital closed n-manifold homeomorphic to the first one. Finally, we obtain six kinds of contractible transformations: deleting a simple point, attaching a simple point, deleting a simple edge, attaching a simple edge, replacing an edge with a point, replacing a simple pair of points with a one point. This paper is organized as follows. In section 2, we give basic definitions related to digital spaces. Different types of contractible digital spaces are defined and studied in section 3. In section 4, we define homotopy equivalence of digital spaces, introduce new types of contractible transformations that can be considered as a digital homeomorphism and present a new proof that the Euler characteristics of homotopy equivalent digital spaces are equal.

## 2. Digital Space General Definitions

As we mentioned above, a digital space can be compared to an atom in which nuclear forces bind nucleons (neutrons and protons) into atomic nuclei.

**Definition 2.1**

A digital space $G=(V,W)$ is a pair of sets $V$ and $W$.
- $V=\{v_1,v_2,...v_n,...\}$ is a finite or countable set of points.
- $W$ is a set of edges. Each edge $(v_p v_q)$ connects two different points, $v_p, v_q \in V$, $v_p \neq v_q$. Two edges $(v_p v_q)$ and $(v_q v_p)$ are the same. Two points can only be connected by one edge: $W=\{(v_p v_q), | v_p, v_q \in V, v_p \neq v_q, (v_p v_q)=(v_q v_p)\}$ (see fig.2.1).

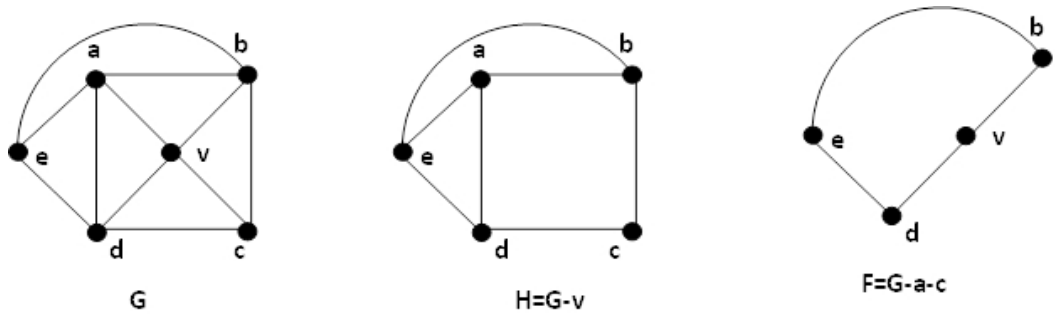

Figure 2.1. G is a digital space. H and F are digital subspaces of G.

Note that digital spaces are called simple undirected graphs in graph theory. We will use the terminology of graph theory whenever it is convenient.

For an edge (uv) of G, the points u and v are called its endpoints and u and v are incident with (uv). Points u and v are called adjacent or neighbors if they are the endpoints of an edge (uv).

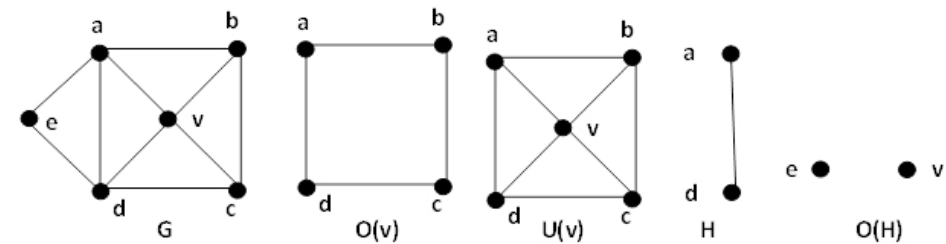

Figure 2.2. G is a space. O(v) is the rim of point v. U(v) is the ball of point v. H is the subspace of G, O(H)={e,v} is the joint rim of H.

We use the notations v∈G and (vu)∈G if v∈V and (vu)∈W respectively, if no confusion can result. Such notions as the connectedness, the adjacency, the dimensionality and the distance on G are completely defined by sets V and W (e.g. [6, 9, 21, 27]).

**Definition 2.2**

The digital space H=(P,S) is called the subspace of the digital space G=(V,W), if P⊆V and H is induced by the set of points P.

Obviously, S is the set of edges that have both endpoints in P. H may be denoted by G[P]. According to this definition, H is obtained by removing from G points that are not contained in P together with their incident edges. In Figure 2.1 the subspace H is obtained by removing the point v from G, the subspace F is obtained by removing points a and c from G.

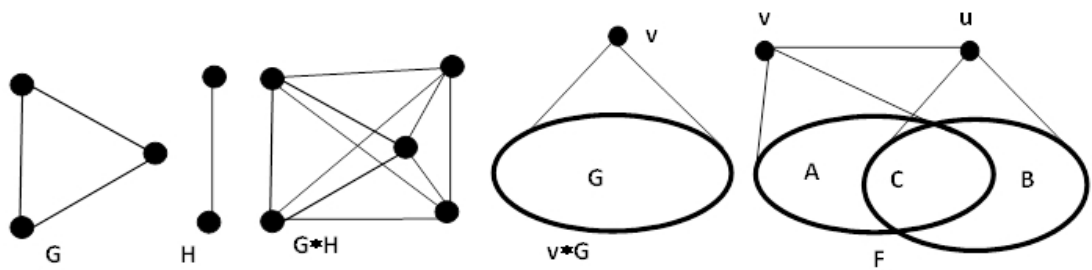

Figure 2.3. G∗H is the join of spaces G and H. v∗G is the cone of G. F is the double cone.

**Definition 2.3**
- Let G and v be a digital space and a point of G. The subspace O(v) induced by the set of points of G that are adjacent to v (without v) is called the nearest neighborhood or the rim of point v in G, (fig. 2.2). The subspace $O(v) \cup v = U(v)$ is called the ball of v.

- Let $G=\{v_1,v_2,v_3,....v_p\}$ and $H=\{v_1,v_2,v_3,....v_n\}$ be a digital space and its subspace. The subspace $O(H)=O(v_1) \cap O(v_2) \cap O(v_3) \cap ... \cap O(v_n)$ is called the joint neighborhood or the joint rim of H (of points $v_1, v_2, v_3,.... v_n$) (fig. 2.2).

In fact, the joint rim O(H) of the collection of points H is the set of all points which are simultaneously adjacent to every single point of H. In fig. 2.2, $O(v)=\{a,b,c,d\}$ is the rim of point v, $U(v)=\{v,a,b,c,d\}$ is the ball of point v, $H=\{a,d\}$ is a subspace of G. $O(H)=\{e,v\}$ is the joint rim of subspace H.

**Definition 2.4**
- A digital space is complete if every pair of distinct points is connected by an edge.
- The join G∗H of spaces G and H with disjoint point sets is the digital space that contains G, H and all edges joining every point in G with every point in H.
- If a point v is adjacent to all points of a digital space G, then the join v∗G is called the cone of G.
- Let a digital space $G=O(v) \cup O(u)$ consists of adjacent points v and u and spaces A, B and C, where the rims $O(v)= A \cup C \cup u$ and $O(u)= B \cup C \cup v$ and any point x of A is not adjacent to any point y of B. We say that the pair {v,u} is **a** simple pair of points and $G=\{v,u\} \cup A \cup B \cup C$ is the double cone.

The join G∗H, the cone of G and the double cone F are displayed in Figure 2.3.

Remind that the digital space K(m) with m points is called complete if every two points of K are adjacent. The path is a digital space P=(V,W) if $V=\{v_1, v_2,... v_n\}$ and $W=\{ (v_i v_{i+1}) \mid i = 1,2,...n-1\}$. Points $v_1$ and $v_n$ are called endpoints of P. A digital space G is called connected if for any two points x and y of G there is a path P with endpoints x and y contained in G, i.e. P is a subspace of G, P⊆G. Notice that we will use some symbols, notations and names introduced in our previous works (e.g. [ 6, 8, 9, 20, 26, 27]).

## 3. Contractible Digital Spaces and Simple Points

Contractible molecular spaces and their intersection graphs were defined and studied in
[ 7, 10, 14-16, 18-31]. In this section we define new types of contractible digital spaces and investigate their properties. To define contractible digital spaces, we use the inductive definition.

**Definition 3.1**

The one-point digital space K(1)=v is the contractible space.
Let G be a contractible space containing n-points, |G|=n and H be a contractible subspace of G, H⊆G. Then the space P=G∪x, that is obtained by attaching a point x to G in such a way that the rim O(x)=H, is a contractible space.

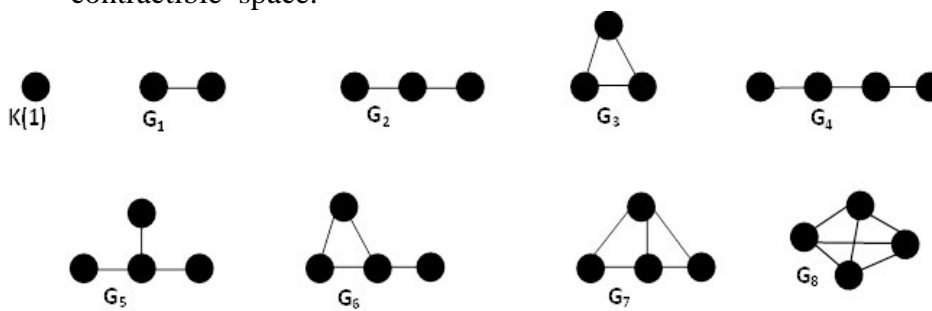

Figure 3.1. Contractible digital spaces with a number of points less or equal to 4.

It is important that the rim O(x) of the point x in P is a contractible space. Denote by T the family of contractible digital spaces. Contractible digital spaces with a number of points less than or equal to 4 are depicted in fig. 3.1.

**Definition 3.2**
- A point v of the digital space G is called simple if the rim O(v) is a contractible space.
- Deleting a simple point x from G means converting G to the subspace G-x without x.
- Attaching a simple point x to G means converting G to the space G∪x, that is obtained by attaching x to G in such a way that the point x is simple in G∪x.

Let us consider properties of contractible digital spaces.

**Proposition 3.1**
A contractible digital space G can be transformed into a point of G by sequentially removing simple points.
Proof.
By definitions 3.1 and 3.2, G is obtained from the one-point space K(1) by sequential attaching simple points. Therefore, G can be converted to K(1) by sequentially removing simple points in the reverse order. The proof is completed. □

**Proposition 3.2**
The complete space K(n) Is contractible.
Proof.
We use induction on the number of points of K(n). It is obviously true when K(n) has a small number n of points, n=1-4 (see fig. 3.1). Assume that it is valid for n≤m and consider K(m). Attach a simple point v to K(m), where O(v)=K(m). By definitions 3.1, K(m)∪v is the contractible space. By construction, K(m)∪v=K(m+1) is the contractible space. The proof is completed. □

**Proposition 3.3**
The join K(1)∗G of the one-point space K(1) and a space G is a contractible space.
Proof.
We use induction on the number of points of G, n=|G|, G= {$x_1, x_2, ..x_n$}, K(1)={v}, K(1)∗G=v∗G. For small n=1÷3, the statement is obvious (see fig. 3.1). Suppose it is true for |G|<m. Let |G|=m, G= {$x_1, x_2, ..x_m$}. The rim of the point $x_1$ in v∗G is O($x_1$)=v∗H, where H is the rim of point $x_1$ in G. O($x_1$)=v∗H is a contractible space by the induction hypothesis. This means that $x_1$ is the simple point in v∗G and can be deleted from v∗G. The space v∗(G-$x_1$) is contractible by induction hypothesis. Therefore, v∗G is contractible by definition 3.1. This completes the proof. □

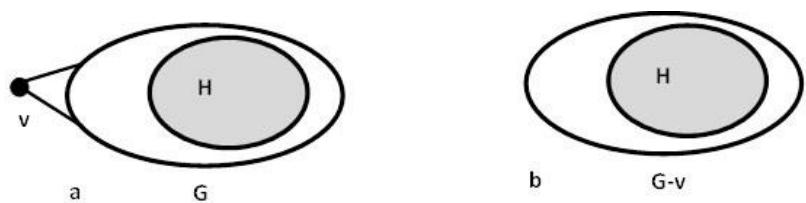

Figure 3.2. a) O(v) is contractible. b) (G-v) is contractible.

**Proposition 3.4**

The join G∗H of a contractible space G and a space H is a contractible space.
Proof.
We will prove this statement by induction on the number of points of the contractible space G. The statement is obvious if G contains one or two points. Suppose it is true for |G|<m. Let |G|=m. Let point x be a simple in G. Then the rim O(x)=A of x in G and the space G-x are both contractible spaces by definition 3.1. By construction, the rim of x in G∗H is A∗H, where |A|<m and (G∗H)-x=(G-x)∗H, where |(G-x)|<m. This means that A∗H and (G∗H)-x are contractible spaces by the induction hypothesis. Therefore, G∗H is contractible by definition 3.1. This completes the proof. □

**Property 3.5**
Any contractible space is connected.
Proof.
Note that a contractible space G can obtained from the one-point space K(1) by sequentially attaching simple points. For a small number n of points it is verified directly (see fig. 3.1). Attach a simple point v to G. Since O(v) is contractible, then O(v) is connected and G∪v is connected by the construction. The proof is completed. □

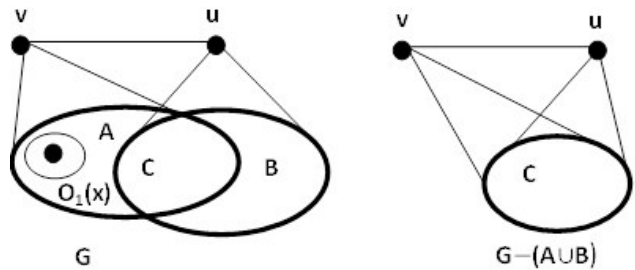

Figure 3.3. G is the double cone. G−(A∪B) is a contractible space.

**Property 3.6**
If H is a contractible subspace of a contractible space G, then H can be obtained from G by sequentially removing simple points.
Proof.
We use induction on the number of points n=|G|. For small n the theorem is verified directly (see fig. 3.1). Assume that the theorem is valid for n≤m. Let |G|=m. Let H be a contractible subspace of a contractible space G. If H contains all simple points, then H=G by definition 3.1. Since H≠ G, then there is a simple point v that does not belong to H, v∈G, v∉H (see fig. 3.2a). Therefore, G-v is a contractible space and H is a subspace of G-v (see fig. 3.2b). By the induction hypothesis, H can be obtained from G-x by sequential deleting of simple points This completes the proof. □

**Proposition 3.7**
The double cone G is a contractible space.
Proof..
Let G=(vu)∪A∪B∪C, where points v and u are adjacent. The rims O(v)= A∪C, O(u)=C∪B and O(v)∩O(u)=C (fig. 3.3). Consider a point x of A, x∈A. The rim O(x)=v∗$O_1$(x), where $O_1$(x)⊆O(v) by construction of G. Hence, x is the simple point and can be deleted from G.

For the same reason, all points belonging to subspaces A and B are simple and can be deleted from G. The obtained space G−(A∪B)=(vu)∗C is contractible by proposition 3.4 . Therefore, G is a contractible space. This completes the proof. □

**4. The Homotopy on Digital Spaces**

Homotopy equivalence of molecular spaces and graphs was defined and studied in [20-23, 25-30]. We define the homotopy on digital spaces in a similar way using attachments and deletions of simple points.

**Definition 4.1**
Digital spaces G and H are called homotopy equivalent if G can be obtained from H by a sequence of attaching and removing simple points. Attaching and removing simple points are called contractible transformations.

Notice that attaching a simple point v to G and removing a simple point v from H=G∪v are mutually inverse operations. The following proposition is the direct consequence of Definition 4.1 and Definition 3.1.

**Proposition 4.1**
All contractible spaces are homotopy equivalent. Any contractible space G can be converted to a point of G by sequentially removing simple points.

Properties of the Euler characteristic and homology groups of molecular spaces and their intersection graphs were studied in [20-23,25, 28-30]. We now define the Euler characteristic of digital spaces as follows:

**Definition 4.2**
Let G be a digital space and $n_p$ be the number of its complete subspaces containing p points. Denote e(G)=$(n_1,n_2,…,n_s)$ as the e-vector of G. The Euler characteristic of G is χ(G)=$n_1-n_2+…+(-1)^{s-1}n_s=\sum_{k=1}^{s}(-1)^{k-1}n_k$.

First, let us study the Euler characteristics of contractible spaces.

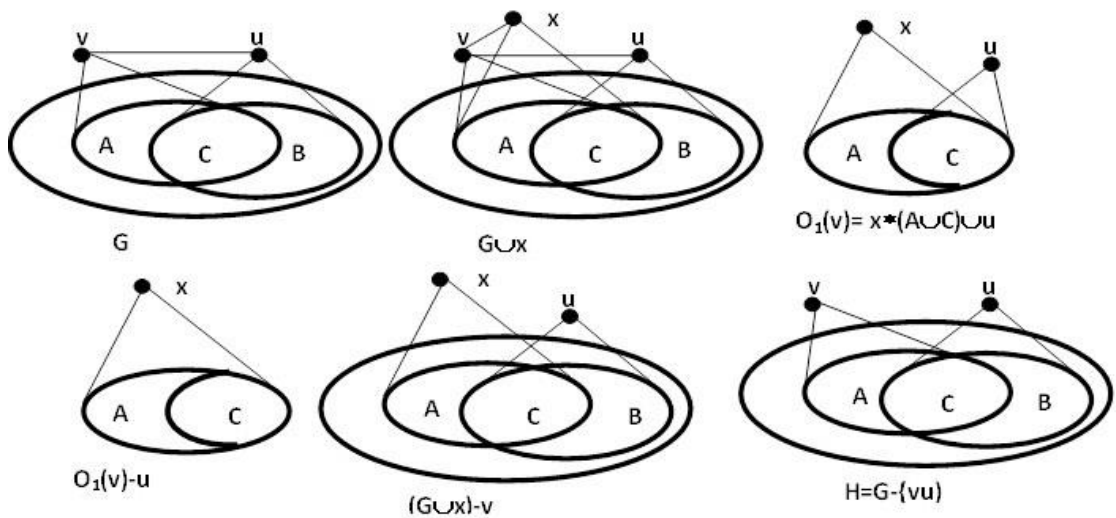

Figure 4.1. {vu} is a simple edge of G. G and H=G-{vu} are homotopy equivalent.

**Proposition 4.2**
If a digital space G is contractible, then the Euler characteristic of G is 1, χ(G)=1.
Proof.
We use induction on the number n=|G| of points of G. It is obviously true when G has a small number n of points, n=1-4 (see fig. 3.1). Assume that it is valid for n≤s and consider G with s points s=|G|. Attach a simple point v to G, where O(v)=H is

a contractible subspace of G. We obtain the contractible space G∪v. By construction, the number |H| of points of H is |H|≤m. Let the vectors e(G) and e(H) be e(G)=($n_1,n_2,…,n_p$) and e(H)=($m_1,m_2,…,m_p$).

It follows from the structure of G∪v that the vector e(G∪v)=($n_1+1, n_2+m_1, n_3+m_2 ,…,n_p+m_{p-1}, m_p$). Then The Euler characteristic of G∪v is χ(G∪v))=($n_1+1$)-($n_2+m_1$)+($n_3+m_2$)-…+$(-1)^{p-1}$ ($n_p+m_{p-1}$)+$(-1)^p$ $m_p$=($n_1-n_2+n_3-…+(-1)^{p-1} n_p$)+1-($m_1-m_2+m_3-…+(-1)^{p-1}$ $m_p$)=χ(G)+1-χ(H). χ(G)=χ(H)=1 by the induction hypothesis. Therefore, χ(G∪v))=1+1-1=1. The proof is completed. □

Consider now the Euler characteristics of homotopy equivalent digital spaces.

**Proposition 4.3**

The Euler characteristics of homotopy equivalent digital spaces G and F are equal, χ(G)=χ(F).

Proof.

Let F is obtained by attaching a point v to G in such a way that the point v is simple in

F=G∪v. This means that the rim O(v)=H is a contractible subspace of G. By definition 4.1, spaces G and F are homotopy equivalent. Using the same reasoning as in Proposition 4.2, we get that χ(F)=χ(G∪v)=χ(G)+1-χ(H). As H is contractible space, then χ(H)=1 by Proposition 4.2. Hence, χ(F)=χ(G). The proof is completed. □

Now we consider several additional operations that transform one digital space into another that is homotopically equivalent to the first space. First let us define simple edges and study their properties.

**Definition 4.4**

- Let points v and u of G are adjacent and the joint rim O(vu)=O(v)∩O(u) is a contractible space. Deleting a simple edge (vu) from G transforms G into the space H=G-(vu) in which points v and u are not adjacent.
- Let points v and u of G are not adjacent and the joint rim O(vu)=O(v)∩O(u) is a contractible space. Attaching a simple edge (vu) to G transforms G into the space H=G∪(vu) in which points v and u are connected by the edge.

**Proposition 4.4**

Let the digital space H=G-(vu) is obtained from a digital space G by deleting the simple edge (vu). Then spaces G and H are homotopy equivalent (see fig. 4.1).

Proof.

Let O(v)= u∪A∪C, O(u)=v∪B∪C and O(vu)=O(v)∩O(u)=C, where C=O(vu) is a contractible space. According to definition 4.1, we need to show that H is obtained from G by attaching and removing simple points. The subspace v∗(A∪C) of G is contractible by property 3.3. Let us attach a simple point x to G so as O(x)=v∗(A∪C).

In the obtained space G∪x, the rim of the point v is $O_1$(v)= x∗(A∪C)∪u. In $O_1$(v), the point u is simple since O(u)=C is a contractible space. Hence, the point u can be deleted and $O_1$(v)-u is a contractible space. Therefore, $O_1$(v) is a contractible space. This means that the point v is simple in G∪x and can be deleted from G∪x. We obtain the space {G∪x}-v, which is isomorphic to H=G-(vu). This completes the proof. □

**Remark 4.1**.

Deleting a simple edge from a digital space and attaching a simple edge to a digital space and are mutually inverse operations. Therefore, digital spaces G and

H are homotopy equivalent if G can be obtained from H by deleting the simple edge (vu)

Let's introduce the replacement of an edge with a point that transforms a digital space G to the homotopy equivalent digital space H. The replacement of an edge with a point increases the number of points in a digital space.

**Definition 4.5**

Replacing an edge with a point in the digital space G consists of attaching a point x to the subspace v∪u∪O(vu) and removing an edge (vu).

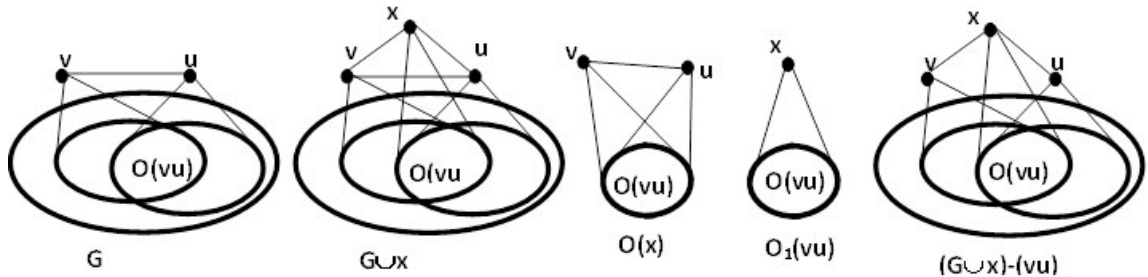

Figure 4.2. Replacing an edge (vu) with a point x in the digital space G.

**Proposition 4.5**

Let G be a digital space and (vu) be an edge in G. Let the digital space H be obtained from G by replacing the edge (vu) with a point x. Then the spaces H=(G-(vu))∪x and G are homotopy equivalent.

Proof.

By the construction of the space G, the subspace v∪u∪O(vu) is contractible (see fig. 4.2). Attach the simple point x to G so that O(x)=v∪u∪O(vu). The edge (vu) in the obtained space G∪x is simple because the joint rim of v and u is the join $O_1(vu)=x*O(vu)$. So the edge (vu) can be deleted from G∪x. The obtained space (G∪x)-(vu) is homotopy equivalent to G. This completes the proof. □

Let us define an operation that reduces the number of digital space points. This can be done if the digital space contains a subspace that is a double cone.

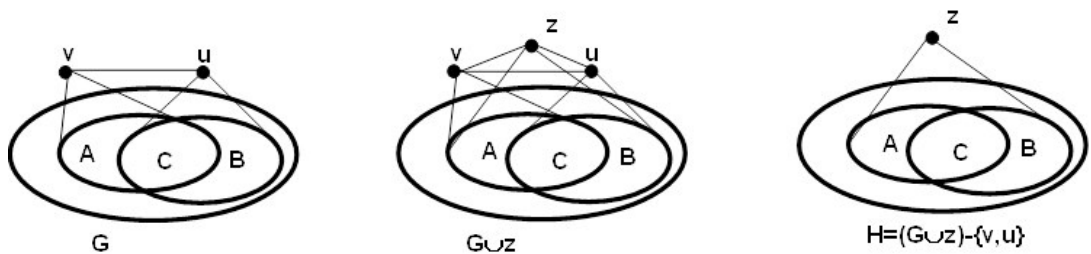

Figure 4.3. {v,u} is a simple pair in G. P=O(v)∪O(u)={v,u}∪A∪B∪C. G and H=G∪z−{v,u} are homotopy equivalent.

**Definition 4.6**

Let a digital space G contains a simple pair of points {v,u}. Attaching a point z to a subspace O(v)∪O(u) and removing points v and u is called replacing a simple pair with a one point.

**Proposition 4.6**

If the digital space H is obtained from the digital space G by replacing a simple pair with a one point, then the spaces H and G are homotopy equivalent.

Proof.

Let {v,u} is a simple pair of points in G. P=O(v)∪O(u) is the subspace of G and the double cone by definition 2.4 (see fig. 2.3). According to proposition 3.7, P is a contractible space. Attach a simple point z to G so that O(z)=P (see fig. 4.3). In the obtained space G∪z, the rim of v is $O_1(v)$=z∗O(v), the rim of u is $O_1(u)$=z∗O(u), i.e., $O_1(v)$ and $O_1(u)$ are contractible spaces. Therefore, points v and u are simple and can be deleted from G∪z. We obtain the space H=(G∪x)-{v,u}. which is transformed from G by attaching and removing simple points. Hence, G and H are homotopy equivalent. This completes the proof. □

Figure 4.3 illustrates the proof of proposition 4.6.

Now we list the operations that transform a digital space into another digital space homotopically equivalent to the first one.
(DSP) Deleting a simple point.
(ASP) Attaching a simple point.
(DSE) Deleting a simple edge.
(ASE) Attaching a simple edge.
(REP) Replacing an edge with a point.
(RSP) Replacing a simple pair of points with a one point.

## 5. Conclusion

This paper introduces new types of contractible spaces. A double cone is defined and it is shown that the double cone is the contractible digital space. On this bases, new types of contractible transformations: the replacement of an edge with a point and the replacement of two points with one point are introduced and studied. These transformations convert a digital closed n-manifold to another digital closed n-manifold homeomorphic to the first one. In fact, these transformations are digital counterparts of a homeomorphism in classical topology. Finally, all contractible transformations are divided into six classes.